%% file: main.tex
\documentclass[10pt,conference]{IEEEtran}
\IEEEoverridecommandlockouts
\usepackage{float}
\usepackage{cite}
\usepackage[T1]{fontenc}
\usepackage{amsmath,amssymb,amsfonts}
\usepackage{algorithmic}
\usepackage{graphicx}
\usepackage{textcomp}
\usepackage{xcolor}
\usepackage{booktabs}
\def\BibTeX{{\rm B\kern-.05em{\sc i\kern-.025em b}\kern-.08em
    T\kern-.1667em\lower.7ex\hbox{E}\kern-.125emX}}

\begin{document}

\title{Prompt-with-Me: in-IDE Structured Prompt Management for LLM-Driven Software Engineering}

\author{\IEEEauthorblockN{Ziyou Li}
\IEEEauthorblockA{
\textit{Delft University of Technology}\\
\textit{JetBrains Research}\\
Delft, The Netherlands \\
ziyou.li@tudelft.nl}
\and
\IEEEauthorblockN{Agnia Sergeyuk}
\IEEEauthorblockA{
\textit{JetBrains Research}\\
Belgrade, Serbia \\
agnia.sergeyuk@jetbrains.com}
\and
\IEEEauthorblockN{Maliheh Izadi}
\IEEEauthorblockA{
\textit{Delft University of Technology}\\
Delft, The Netherlands \\
m.izadi@tudelft.nl}
}

\maketitle

\begin{abstract}

Large Language Models are transforming software engineering, yet prompt management in practice remains ad hoc, hindering reliability, reuse, and integration into industrial workflows. We present \textit{Prompt-with-Me}, a practical solution for structured prompt management embedded directly in the development environment. The system automatically classifies prompts using a four-dimensional taxonomy encompassing intent, author role, software development lifecycle stage, and prompt type. To enhance prompt reuse and quality, Prompt-with-Me suggests language refinements, masks sensitive information, and extracts reusable templates from a developer’s prompt library.

Our taxonomy study of 1,108 real-world prompts demonstrates that modern LLMs can accurately classify software engineering prompts. Furthermore, our user study with 11 participants shows strong developer acceptance, with high usability (Mean SUS=73), low cognitive load (Mean NASA-TLX=21), and reported gains in prompt quality and efficiency through reduced repetitive effort. 
Lastly, we offer actionable insights for building the next generation of prompt management and maintenance tools for software engineering workflows.

\end{abstract}


\section{Introduction}

Large Language Models (LLMs) such as GPT‑4 and Claude are rapidly transforming modern software engineering (SE). Tools powered by these models have moved from experimental prototypes to integral components of everyday development workflows. For instance, industry reports on GitHub Copilot, a GPT‑powered AI pair programmer, show that developers using the tool complete coding tasks up to 55\% faster, while 85\% report increased confidence in their code quality~\cite{research_research_2024}. These numbers highlight not only the efficiency gains brought by LLMs but also their growing influence on developer experience and software quality.

Despite the transformative role of LLMs in software engineering, the primary interface for human‑LLM interaction, the \textit{prompt}, remains surprisingly informal. Prompts are typically written ad hoc, often with fragmented grammar or inconsistent wording, even though minor phrasing changes can significantly alter model behavior\cite{chatterjee_posix_2024}. Unlike source code, prompts are rarely versioned, reviewed, or maintained across the software development lifecycle, leaving critical interactions with LLMs effectively unmanaged. A recent study of 243 GitHub repositories found that prompt modifications are sparsely documented and frequently introduce logical inconsistencies~\cite{tafreshipour_prompting_2025}. As LLMs become embedded in production workflows, software teams face an emerging challenge: how to systematically treat prompts as first‑class software artifacts.

To address these challenges, we offer \textit{prompt maintenance}: the ongoing curation, documentation, and evolution of a team’s prompt library to support everyday software development. This is distinct from prompt \textit{optimization}, which targets improving a single prompt for a specific output metric. Maintenance is crucial because prompts serve diverse purposes, must adapt as requirements evolve, and should remain understandable to humans. Existing tools, however, address only fragments of this need. Automated optimizers, leveraging evolutionary algorithms or reinforcement learning, can mutate prompts to improve performance~\cite{wen_hpss_2025, agarwal_promptwizard_2024}, yet often produce ambiguous text that still requires human validation~\cite{zhou_revisiting_2023}. Meanwhile, LLM‑Ops platforms such as PromptLayer and Helicone~\cite{magnivorg_promptlayer_2025, helicone_heliconehelicone_2025} operate outside the Integrated Development Environment (IDE) and are not tailored to prompts used during active software development. Existing IDE plugins offer only prompt storage~\cite{jetbrains_prompt_2025}, providing no support for quality assurance, duplicate detection, or developer comprehension. Consequently, the software engineering community currently lacks an integrated solution for maintaining programming prompts throughout the entire software development lifecycle.

In this paper, we present \textbf{Prompt‑with‑Me}, a JetBrains IDE plugin designed to bridge the gap in prompt maintenance for software engineering. The tool integrates prompt management directly into the developer workflow, offering a searchable prompt library enriched with automated quality and optimization features. Each new or modified prompt is automatically classified using a research‑driven taxonomy that captures its \textit{intent}, \textit{author role}, \textit{SDLC phase}, and prompt \textit{type}. Prompt‑with‑Me also provides real‑time feedback, highlighting spelling or grammar issues, potential sensitive data, and opportunities for correction. To reduce redundancy, the tool detects duplicate prompts and, when confirmed, extracts parameterized templates for reuse. All these processes are delivered within the IDE through a guided, developer‑friendly interface. This allows users to accept, edit, or ignore optimization suggestions without disrupting their workflow.

To evaluate Prompt‑with‑Me, we formulate the following four research questions (RQs):
\begin{itemize}
     \item \textbf{RQ1}: To what extent can LLMs reliably \textit{categorize} prompts used in software engineering workflows?
     \item \textbf{RQ2}: How \textit{usable} is Prompt‑with‑Me as a system for maintaining prompts in software engineering?
     \item \textbf{RQ3}: How \textit{helpful} is Prompt‑with‑Me for users when maintaining prompts within their working environment?
     \item \textbf{RQ4}: What is the impact of using Prompt‑with‑Me on a programmer’s \textit{perceived cognitive load} during prompt maintenance?
\end{itemize}

We address RQ1 through a three‑step process. First, we collected a corpus of 1,108 real‑world SE prompts. Next, we derived a four‑dimensional taxonomy from this corpus, grounded in existing literature and validated by SE experts~\cite{shah_using_2023}. Finally, we applied LLMs as classifiers using few‑shot learning to automatically assign taxonomy labels, achieving substantial inter‑rater agreement with human raters (Fleiss' \(\kappa\)=0.72). 

To investigate RQ2–RQ3-RQ4, we conducted a task‑based moderated user study with 11 software developers from diverse business domains. Prompt‑with‑Me achieved a System Usability Scale (SUS) score of 73/100, indicating good usability. Participants reported that the tool improved prompt quality, reduced repetitive effort, and simplified their workflow compared to existing practices. They further noted that adding, refining, templating, and editing prompts required only light cognitive effort. This suggests that Prompt‑with‑Me effectively supports prompt maintenance without imposing significant mental load.

Our work positions prompts as first‑class software artifacts, opening the door to versioning, quality assurance, and CI/CD integration of prompts. 
Our findings also highlight opportunities for collaborative prompt repositories, refactoring strategies, and scalable prompt analytics, ultimately reducing developer effort and improving the reliability of AI‑assisted software engineering.
In summary, our contributions are as follows:  
\begin{enumerate}
    \item A four‑dimensional taxonomy of SE prompts and an annotated dataset of 1,108 prompts,
    \item Prompt‑with‑Me, to the best of our knowledge, the first IDE‑native tool that unifies prompt storage, classification, optimization, and template extraction,
    \item An empirical evaluation of LLM performance on SE prompt classification,
    \item A user study on system usability, helpfulness, and cognitive load of Prompt-with-Me,
    \item Actionable insights for designing future IDE-native prompt maintenance tools.
\end{enumerate}

\section{Related Works}

The rapid advancement of LLMs has introduced a new paradigm in software engineering, where natural language intent can be translated into executable behavior~\cite{deljouyi2024leveraging,izadi_language_2024,ionescu2025multi}. This shift has spurred the development of tools and frameworks aimed at prompt management and optimization. However, most existing solutions remain disconnected from real-world software engineering practices, treating prompts as short-lived strings rather than persistent, maintainable artifacts within the developer workflow. In this section, we survey prior efforts across two key dimensions, prompt management and prompt optimization, and identify critical gaps that hinder their industrial adoption.

\subsection{Prompt Management Tools}
Artifact visibility and management are crucial in industrial settings.
Software such as PromptLayer~\cite{magnivorg_promptlayer_2025} and Helicone~\cite{helicone_heliconehelicone_2025}  provide dashboards for prompt change tracking.
But they are isolated tools and remain disconnected from the SE toolset.

Some prompt management utilities live inside the IDE itself, narrowing the gap between prompt work and everyday coding.
Azure Prompt Flow~\cite{lgayhardt_prompt_nodate}, a prompts-ops IDE plugin, integrates prompt management.
However, it is targeted for LLM-application building and is not used for direct LLM interaction.
Existing JetBrains AI Assistant Prompt library allows developers to store prompts they know would be reused~\cite{jetbrains_prompt_2025} but offer no optimization mechanism or guidance.
Our system bridges this tooling gap by enhancing JetBrains AI Assistant Prompt library with guided maintenance steps in the workflow that let developers keep, evolve, and review prompts along with other project artifacts.
This operationalizes the methodology proposed by earlier studies and enables real-world evaluation of the methodology~\cite{chen_promptware_2025}.

\subsection{Prompt Improvement}
Systems for prompt improvement often emphasizes on their ability to automate the search or generation of effective prompts for certain tasks.
They treat the search for better prompts as a black-box optimisation problem and apply evolutionary algorithms, Monte-Carlo tree search, Bayesian bandit, or reinforcement learning to mutate candidate prompts~\cite{wen_hpss_2025, murthy_promptomatix_2025, wu_prompt_2024, agarwal_promptwizard_2024}.
However, studies also show that automated prompt optimization does not always outperform carefully crafted manual prompts~\cite{zhou_revisiting_2023}.
In some applications, such as data labeling, automatic pipelines can be unreliable without human oversight~\cite{he_prompting_2025}.
Empirically, this is also the case for the diverse tasks in programming.
These outcomes motivate the addition of human judgment to the optimization loop.
Researchers have started placing users directly into the optimization loop.
\textit{Prompt Optimization with Human Feedback} frames prompt tuning as reinforcement learning with human preference signals~\cite{lin_prompt_2024}.
Interactive systems, such as \textit{Interactive Prompt Optimization}, leverage LLM for candidate prompt generation and present candidate prompts to their user~\cite{li_iprop_2025}.
Also, \textit{Human-in-the-loop LLM-based Agents framework} puts the agent directly in collaboration with LLM-based Agents coding workflow~\cite{takerngsaksiri_human---loop_2025}.
In addition to optimizing prompts directly, a study finds that  well‑structured prompt templates can significantly elevate. instruction‑following performance~\cite{liu_prompt_2024}.
Recent work underscores the ad-hoc nature of prompt template management \cite{mao_prompts_2025}. 
Prompt-with-Me fills these gaps by embedding prompt improvement procedures directly within the JetBrains IDE, supporting automated template extraction, versioning, and categorization. 
This integration reduces cognitive load by aligning prompt management closely with developers' existing workflows.

\section{Proposed Approach}

\begin{figure*}[tb]
\centerline{\includegraphics[width=0.9\textwidth]{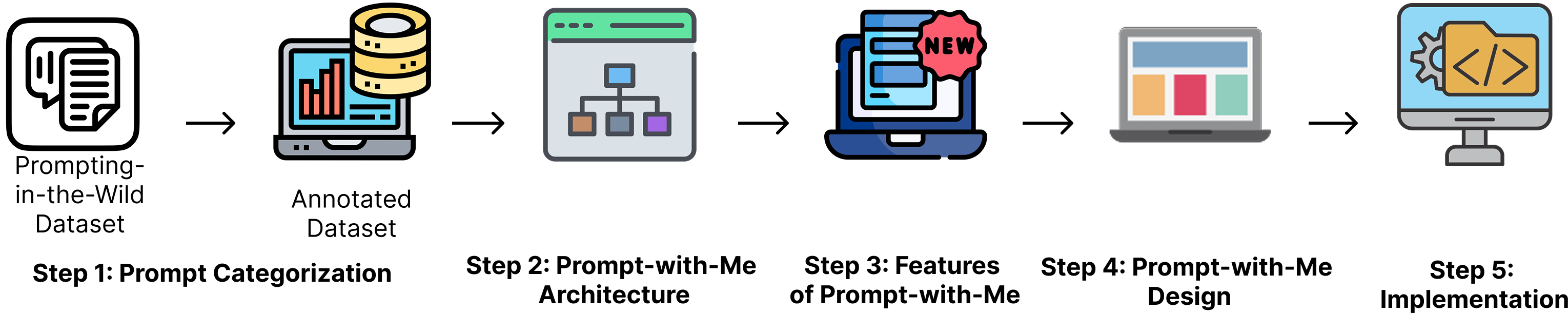}}
\caption{Proposed Approach Structure}
\label{fig:methodology}
\end{figure*}

We introduce Prompt‑with‑Me, an IntelliJ IDEA plugin that treats prompts as first‑class artifacts within the development environment. The plugin provides a structured prompt library and integrates automated tools for prompt categorization and optimization, enabling developers to manage and refine prompts directly within their project context.

Our approach begins with an empirical study of real‑world prompts from software projects to derive a robust categorization scheme, which forms the foundation for clear and maintainable prompt management. Building on this categorization, we design an architecture that incorporates features to organize and optimize the prompts. We then present the core components of this architecture, followed by details of the system’s design and implementation.

The remainder of this section is organized into sections corresponding to these steps: prompt categorization, system architecture, prompt processing, design, and implementation.

\subsection{Prompt Categorization} 

\begin{figure*}[tb]
\centerline{\includegraphics[width=0.9\textwidth]{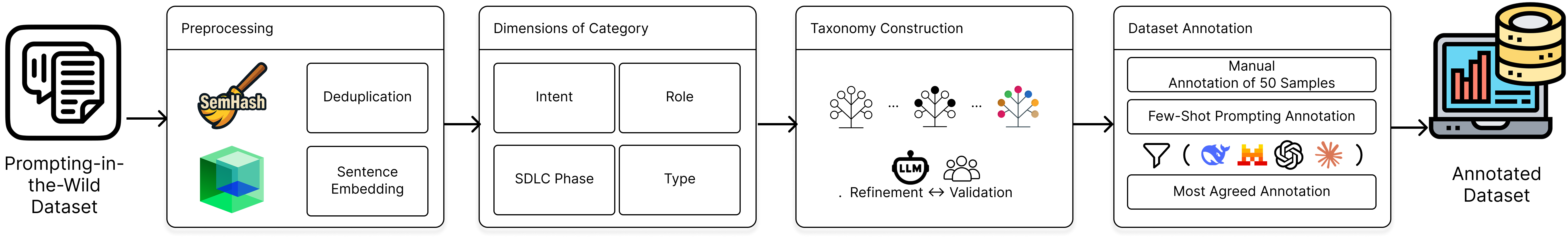}}
\label{fig:prompt_categorization}
\caption{Overview of Prompt Categorization}
\end{figure*}

Understanding the current state of prompts being used is crucial for building the prompt management system.
And with a vision to maintain prompts in a targeted manner, such as applying a certain sequence of maintenance processes to the prompt, we need to construct a categorization of prompts used in SE.
To do this, we employ the Prompting-in-the-Wild dataset~\cite{tafreshipour_prompting_2025}.
The dataset provides 1262 prompt change records across 243 GitHub repositories.
We employ this dataset because it is, by nature, within the context of SE, and is manually maintained by developers in sizable open-source software projects.
In addition, it can be organized into per-repo prompt libraries.
This aligns naturally with the prompt library form of the prompt optimization system and allows further investigation of prompt libraries.

\subsubsection{Preprocessing}
Upon manual inspection, we found that several prompt changes recorded in the Prompting-in-the-Wild dataset are duplicative. 
That is to say, we found that there are multiple instances of extremely similar changes recorded as different diffs in the dataset.
While this may be vital for other studies, such as studies on duplicative commits, we only focus on tracking the prompts themselves.
Therefore, deduplicating these repetitive commits is the most vital.
We used Semhash, a library for text deduplication~\cite{thomas_van_dongen_and_stephan_tulkens_semhash_2025}.
The threshold parameter for deduplication is determined to \(0.999\) through trial and error, through manual verification of the deduplicated prompts by the first author and by an experienced researcher in software engineering with more than 10 years of experience. 
The dataset size is reduced from 1262 to 1108 after deduplication.
Each prompt after deduplication is also embedded using sentence embedder \textit{ibm-granite/granite-embedding-125m-english}.

\subsubsection{Dimensions of Taxonomy}
\label{sec:dim_of_cat}
As the first step in taxonomy construction, we defined four orthogonal dimensions of taxonomy that would be of interest to the prompt optimization system, namely prompt \textbf{intent}, \textbf{SDLC} phase, prompt author \textbf{role}, and prompt \textbf{type}.
Respectively, these dimensions capture the why, when, who and how of a prompt.
These dimensions collectively ground the processing steps we introduce and provide avenues for future work.
The intent dimension captures the prompts’ underlying goal or information need.
The role dimension reflects the professional role of the prompt’s author in the software project (e.g., developer, data scientist, manager).
The SDLC Phase dimension tags each prompt to a stage of the software development life cycle (e.g. design, implementation, testing).
Finally, the dimension of prompt type denotes the style of prompt formulation (for instance, a few-shot example-based query or a template-based prompt with variables).

\input{tables/annotation_results}

\subsubsection{Taxonomy Construction}
\label{sec:taxonomy}
Next, we constructed the taxonomy in the four dimensions.
We employ a method for the construction of taxonomy and the annotation of the dataset in a collaborative human-LLM manner~\cite{shah_using_2023}.
To ensure alignment of the taxonomy content between in-company knowledge and open-source datasets, we first gathered a prompt dataset from internal projects and consented developers.
Then, an initial taxonomy is constructed with the internal dataset.
The taxonomy is validated by an experienced researcher in software engineering with more than 10 years of experience.
Next, we refine the taxonomy in the open-source dataset by adding, removing, merging, or editing the definition of categories through the human-LLM iterative refinement loop.
Finally, this loop ends with the validation of the taxonomy.
The first author validates by checking the annotation of 100 random samples of the open-source dataset.
If fewer than 5 prompts are incorrectly annotated, the taxonomy could be deemed valid and accurate, as provided by the literature~\cite{shah_using_2023}.
The final taxonomy can be found in the Category column of Table~\ref{tab:mistral_comparison}.

\subsubsection{Dataset Annotation}
\label{sec:dataset}
As the data set we used is not annotated, we annotated the data set using the constructed taxonomy in Section~\ref{sec:taxonomy}.

The first author manually annotated a subset of prompts (\(N=50\)).
The remaining prompts were annotated using an optimized hybrid strategy using LLMs with enhanced few-shot prompting techniques.
Each prompt for annotation employed six random manually labeled examples combined with decision trees and chain-of-thought reasoning.
Each few-shot example included both old and new prompt pairs with their complete four-dimensional classifications.

To choose the lease biased annotator LLM, we follow the method of picking the model whose inclusion boosts the agreement the most, as suggested by the study~\cite{davoudi_collective_2025}.
To this end, we employ a \textit{leave-one-out} strategy.
First, we compute \(K_4\), the Fleiss' \(\kappa\) for the four models.
Next, for each model \(i\), we calculate \(K_{3,\overline{i}}\) = Fleiss’ \(\kappa\) on the other three.
Finally, we derive which model's presence causes the largest growth from \(K_{3,\overline{i}}\) to \(K_4\), which can be interpreted as the model with the most contribution, hence the best model.
We experimented with the annotation on four widely-used models (GPT-4o-mini, Claude-3-Haiku, DeepSeek-Chat, and Mistral-Small) via their official APIs with default parameters for each annotation category and for both old and new prompts.
The result could be inferred from Table~\ref{tab:leave_one_out_fleiss}.
We select Mistral as the annotator because it has the highest count of contributions for every combination of old/new prompt and taxonomy dimension.
The distribution of classes of the resulting dataset annotated with the Mistral-Small model can be found in Table~\ref{tab:mistral_comparison}.

\subsection{Prompt-with-Me Architecture Overview}
\begin{figure}[tb]
\centerline{\includegraphics[width=0.95\columnwidth]{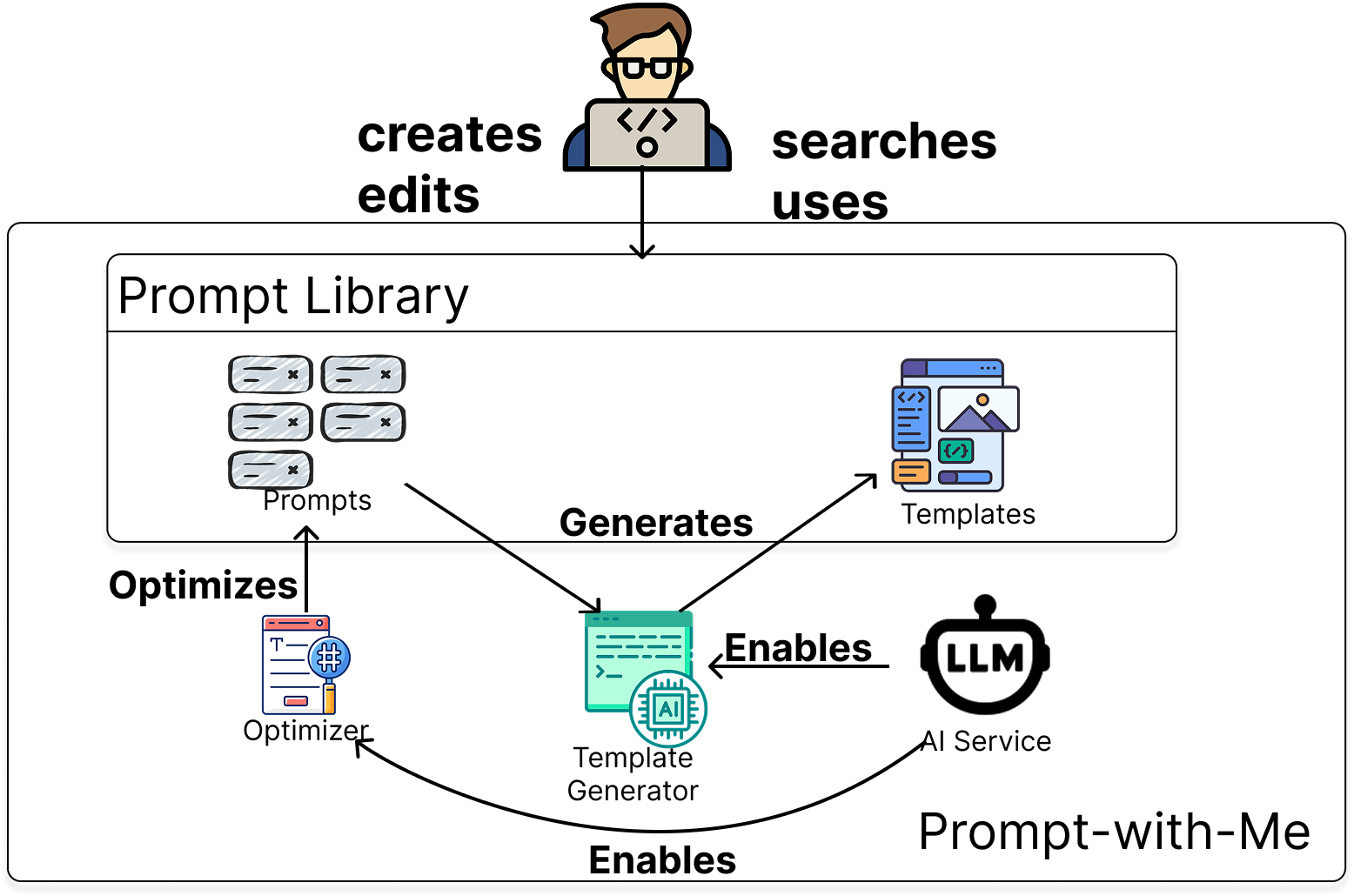}}
\caption{Prompt-with-Me Architecture Overview}
\label{fig:archi_overview}
\end{figure}

The Prompt-with-Me system is organized around the prompt library that stores a set of prompts and templates, as illustrated in Figure~\ref{fig:archi_overview}.
The prompt library allows developers to create, edit, search, and use prompts and templates for their interaction with LLM-enabled applications.
To support the refinement of artifacts in the library, we define two back-end services.
The \textbf{Optimizor} monitors newly added or modified prompts and suggests optimizations to the prompts (e.g., language improvement, anonymisation).
Complementarily, the \textbf{Template Generator} analyses the prompts inside the library.
It proposes a prompt template for the prompts to be stored in the prompt library when certain criteria are met.
In our case of implementation, it is set whenever two prompts exceed a similarity score of 70\%.
Prompt templates are skeletons of many similar prompts, with variables for the developer to fill in for the specific tasks.
Both services call external LLM APIs to perform fine‑grained tasks, such as analyzing prompt wording or extracting template variable that are needed to generate their recommendations.

\subsection{Prompt Processing Steps}
Prompt-with-Me includes a suite of processing steps that effectively manage prompts in software development.

\subsubsection{Classification}

Prompt-with-Me annotates each newly created or modified prompt along four categorical dimensions, as defined in Section~\ref{sec:dim_of_cat}. These annotations serve as input features for downstream optimization tasks.
To evaluate the feasibility of automatically classifying these dimensions, we experimented with four widely used machine learning models: Random Forest, Logistic Regression, Support Vector Machine, and a three-layer Multi-Layer Perceptron (MLP). All models were implemented using the Scikit-learn library with default hyperparameter settings and were trained and evaluated on our annotated dataset described in Section~\ref{sec:dataset}.
We report the classification performance of each model, measured by the weighted F1-score, in Table~\ref{tab:f1_weighted_comparison}. Among the tested models, the MLP achieved the highest overall classification performance across most dimensions. An exception was observed for prompt type classification, where the Random Forest model yielded superior accuracy. Based on these results, we adopted a hybrid approach: Random Forest is used for prompt type classification, while MLP is employed for the remaining three dimensions.

\begin{table}[tb]
  \centering
    \caption{F1-Weighted Scores of Classifiers across Prompt Categorization Dimensions}
  \label{tab:f1_weighted_comparison}
  \begin{tabular}{l c c c c}
    \toprule
    \textbf{Target} & \textbf{RF} & \textbf{LR} & \textbf{SVM} & \textbf{MLP} \\
    \midrule
    Intent & 0.43 & 0.32 & 0.33 & \textbf{0.45} \\
    Type   & \textbf{0.73} & 0.66 & 0.66 & 0.70 \\
    Role   & 0.75 & 0.64 & 0.71 & \textbf{0.77} \\
    SDLC   & 0.44 & 0.40 & 0.40 & \textbf{0.44} \\
    \bottomrule
  \end{tabular}
\end{table}

\subsubsection{Similarity Detection}
\label{sec:similarity_detection}
To maintain a concise library and encourage the reuse of prompt templates, we implemented a similarity detection system that identifies duplicate or near-duplicate prompts before they are added to the repository. This approach reduces redundancy and ensures that the collection remains both compact and versatile. To achieve robust and reliable detection, we employ a strategy that integrates complementary similarity measures:
\begin{itemize}
    \item \textbf{Levenshtein Distance Similarity:} Calculates edit distance between prompts, normalized to a scale of 0-1.
    \item \textbf{Jaccard Similarity:} Compares word sets between prompts to identify common vocabulary.
    \item \textbf{Cosine Similarity:} Uses character n-grams to detect similar patterns in the prompt.
    \item Produces an ensemble similarity score using a weighted average (40\% Levenshtein, 30\% Jaccard, 30\% Cosine), prioritizing Levenshtein to account for the frequent and semantically meaningful minor edits observed in prompt design, as supported by findings in industrial text similarity studies~\cite{andrzejewski_text_2023}.
\end{itemize}
This ensemble score balances sensitivity to small edits with the ability to capture broader lexical and structural overlaps, thereby providing a comprehensive similarity assessment for prompt management.

\subsubsection{Language Improvement}
The language improvement module focuses on improving prompt clarity and correctness by performing automated spelling and grammar checks. Suggested corrections are presented to users, who can selectively apply them to refine their prompts. The system implements two key language improvements:
\begin{itemize}
    \item \textbf{Spelling Correction:} Spelling errors are identified and corrected using \textit{LanguageTool}. 
    Corrections preserve the original case, and confidence scores are computed based on word length and the significance of the modification.
    \item \textbf{Grammar Refinement:} Grammar issues are detected with LanguageTool and presented as suggestions. 
    These carry slightly lower confidence than spelling corrections, reflecting the inherently more subjective nature of grammar adjustments.
\end{itemize}

\subsubsection{Anonymization}
Prompt-with-Me includes an anonymization feature to protect sensitive information in prompts.
The system uses an anonymization service with a local Name Entity Recognition (NER) model and redacts information identified as sensitive entities, such as email addresses, API keys, passwords, phone numbers, credit card numbers, IP addresses, URLs, and other personal identifiers to ensure privacy.
When sensitive information is detected, the information is replaced with ``[REDACTED]'' placeholders. 
Anonymization suggestions are presented with high confidence (0.95-0.99) due to their importance for security and privacy, to raise awareness among users.

\subsubsection{Template Generation}
The template generation process transforms the prompts in the library into reusable templates with variable sets. 
This process can be triggered manually or automatically when multiple similar prompts are detected with the similarity detection feature as described in Section~\ref{sec:similarity_detection}. 
The process constructs a template generation prompt, including the prompt selected for template generation, along with similar prompts in the prompt library, and the four-dimensional classification of each prompt to better capture the intent and the needs of different prompt users.
The template generation prompt is sent to the LLM service to analyze text patterns.
The LLM identifies common structures, variable parts, and appropriate variable names based on content analysis.
The service returns a JSON response containing the generated template with placeholders (e.g., {{language}}, {{value}}), identified variables, and confidence scores for each suggestion.
This approach ensures templates are contextually aware and maintain the original prompt's intent while providing flexibility through well-named variables.

\subsubsection{Summarization}
Prompt-with-Me provides library summarization capabilities to help users understand the overall content and patterns in their prompt collections.
The summarization process analyzes the entire prompt library to extract key insights:
\begin{itemize}
    \item \textbf{Topic Analysis: }LLM service identifies the main topics covered by prompts in the library.
    \item \textbf{Intent Distribution: }Determines the most common intent categories.
    \item \textbf{Role Targeting: }Identifies which professional roles the prompts are primarily designed for.
    \item \textbf{TL; DR\footnote{Short for ``too long; didn't read''.}: }50 to 100 words generated from the library content describing overall purpose of the library.
\end{itemize}

The summary provides a high-level overview of the prompt library's focus and usage patterns. 
This feature is particularly valuable for teams managing prompt libraries, as it provides quick insights without review of all prompts.

\subsection{Prompt-with-Me Design}
\begin{figure}[tb]
\centerline{\includegraphics[width=\columnwidth]{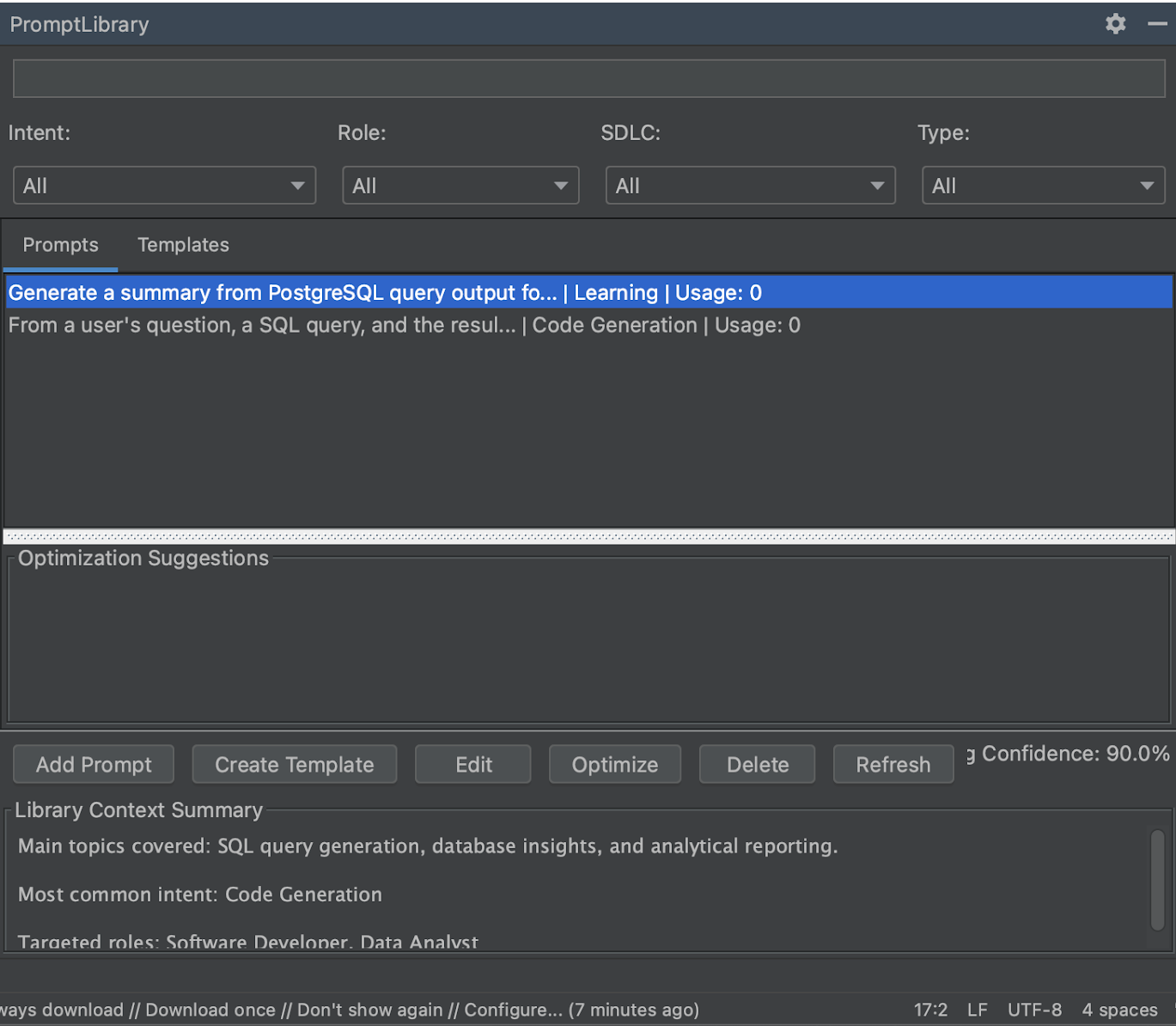}}
\caption{Prompt-with-Me Plugin Window in JetBrains IntelliJ IDEA IDE}
\label{fig:pwm_window}
\end{figure}

Prompt-with-Me is presented as a dedicated tool window, as shown in Figure~\ref{fig:pwm_window}.  
A filterable master list occupies the top half, while a detail pane below streams optimization suggestions for the prompts in the library.  
The tabs switch between prompts and templates, and the header dropdowns filter by intent, role, SDLC, and type.  
A toolbar offers Add, Template, Edit, Optimize, Delete, and Refresh actions.
Although the current design requires further refinement, this comprehensive layout incorporates all the actions and displays required by the developer.

\subsection{Implementation Details}
\begin{figure}[tb]
\centerline{\includegraphics[width=\columnwidth]{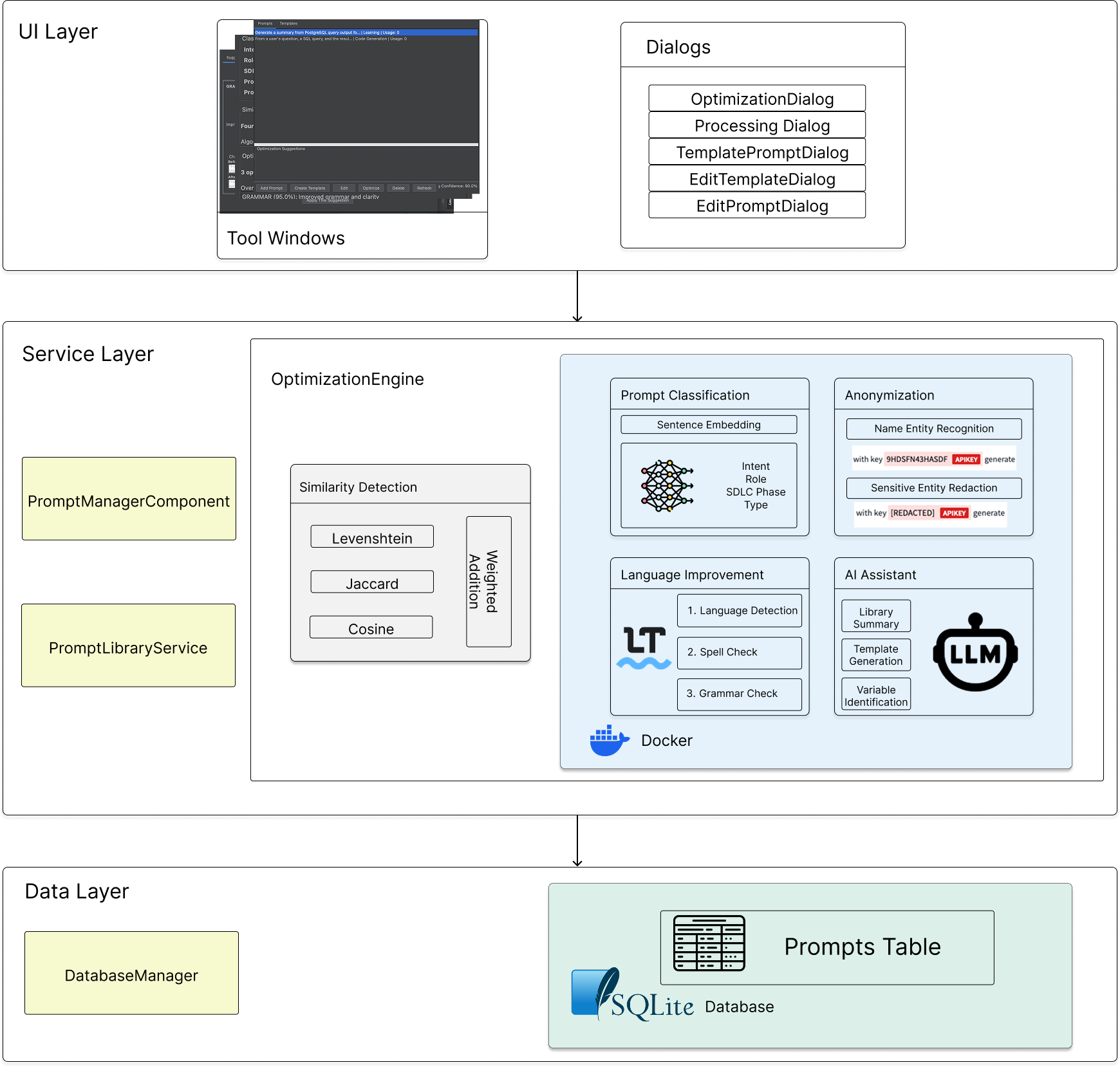}}
\caption{Implementation Diagram}
\label{fig:implementation}
\end{figure}

The Prompt-with-Me system is implemented as a JetBrains IDE plugin using Kotlin, with a layered architecture that separates UI, service, and data concerns.
An overview of the implementation can be found in Figure~\ref{fig:implementation}.
The UI layer includes tool windows and dialogs for prompt management.
The service layer contains core business logic through services.
The data layer uses SQLite for local storage through a DatabaseManager.
Several advanced capabilities are implemented as Docker services, including prompt classification, anonymization, language improvement, and a general AI assistant for template generation and summarization using the DeepSeek API.
The plugin communicates with these services via HTTP APIs.

\section{User Study}
\begin{figure*}[tb]
\centerline{\includegraphics[width=0.95\textwidth]{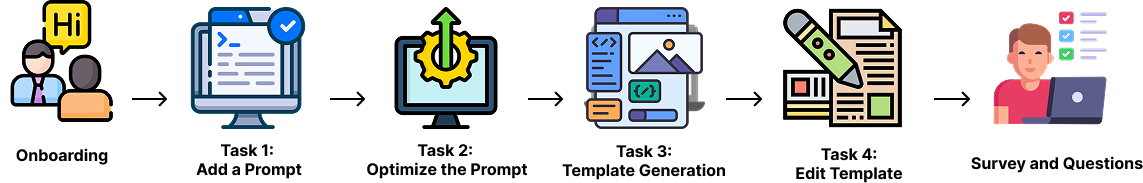}}
\caption{User Study Workflow}
\label{fig:user_study}
\end{figure*}

\subsection{Study design}
Our goal for the user study is to investigate the usability, the helpfulness, and the impact on perceived cognitive load of Prompt-with-Me in a realistic SE context.
To test Prompt-with-Me in a realistic SE context, we needed an initial prompt library to act as initial data in Prompt-with-Me.
Therefore, we decided to choose one repository from the dataset built earlier in Section~\ref{sec:dataset} as the prompt library for the user study case.

To find the appropriate repository in the dataset, we decided on several criteria to filter.
An eligible repository should include at least two SE-related prompts (not annotated as "General") for the SDLC and role category.
The repository should have at least 2 commits for prompts.
In addition, there should at least be 2 files.

Among the repositories eligible for these criteria, we manually investigated the prompts and decided to use the prompts in repository \textit{Dataherald/dataherald} for the following reasons.
First, the intents of the prompts are varied, including Code Generation, Documentation \& Explanation, and Learning.
Second, the general scenario of the prompts are SQL querying, which is a common application scenario in organizational software development, as reported by Stack Overflow~\cite{stack_overflow_stack_2023}.
Third, the repository records explicit changes in prompt type between older and newer prompts.
Specifically, prompt type change from \textit{Zero-shot} to \textit{Template-based}.
This suggests that the prompt's capability of being modified into a template is available, which is a fit for testing the template generation ability of Prompt-with-Me.

Based on the selected prompts in the repository, we design a case study that allows testing of the main functionalities of Prompt-with-Me. 
The case study replicates a realistic software development context where participants are asked to manage prompts used in LLM-assisted SQL querying tasks.

The study followed these steps:

\textbf{(1) Prelude: }
We first introduced the study's goals and procedures, obtained informed consent, and asked participants background questions about their occupation, coding experience, and how often they use LLMs at work.

\textbf{(2) Onboarding: }
The participants are asked to assume the role of a software developer working at a data analytics company.
Their primary tasks involve writing, editing, and optimizing prompts that are used to interact with LLMs to generate or summarize SQL queries.
Although the technical requirement for tasks is intentionally abstracted away, the scenario remains grounded in tasks that reflect actual prompt maintenance behaviors in enterprise workflows.

\textbf{(3) Main Study: } 
Participants start JetBrains IntelliJ IDEA IDE with the Prompt-with-Me plugin, in which a prompt library containing two pre-existing prompts related to SQL summarization and response generation exists. 
They are then asked to perform the four tasks using Prompt-with-Me. 
First, they add a new prompt to the library and observe how the system gives basic properties of the prompts (e.g., length, word count), automatically classifies it in the four dimensions defined in Section~\ref{sec:dim_of_cat}, and suggests optimizations suitable for the prompt.
Next, they initiate the optimization process for this prompt, accepting suggestions related to language improvement and anonymization. 
In the third task, participants are asked to generate a prompt template from the prompt added in the first task.
They are requested to verify whether the automatically extracted variables for the template are reasonable.
Finally, they edit the generated template to customize it with additional variables and variable sets.

\textbf{(4) Post-Study Survey and Feedback Collection}
After the main study, the participants are asked to evaluate the usability of the system by completing the standard SUS questions~\cite{john_brooke_sus_1996} and to evaluate the perceived cognitive load with NASA Task Load Index~\cite{hart_development_1988}.
Raw NASA Task Load Index is selected as suggested by previous research for early studies~\cite{bustamante_measurement_2008}.
They were also asked the following open questions:
(1) Did Prompt-with-Me help you write prompts more efficiently? If yes, please elaborate how.
(2) What feature(s) or aspect(s) did you find most useful in Prompt-with-Me?
(3) Do you have any suggestions for making Prompt-with-Me more suitable for your needs at work?

During the survey, participants were encouraged to think aloud, and the surveyor took notes and used a transcriber to capture the words of the participants when completing the tasks.
Upon completing the survey, the surveyor asked participants to additionally provide some last comments about the workflow, the task, and our system in relevance of the respective experience of the participants at work.

\subsection{Participants}
The targeted user group of Prompt-with-Me is a group of software professionals working on software projects using JetBrains IDEs.
They interact with LLM-powered tools in their work and would benefit from maintaining a shared prompt library, especially in settings where tasks have repeatable structures with minor variations.
This situation with significant context reliance makes prompt maintenance valuable for reuse and adaptation.

\begin{figure}[tb]
\centerline{\includegraphics[width=0.85\columnwidth]{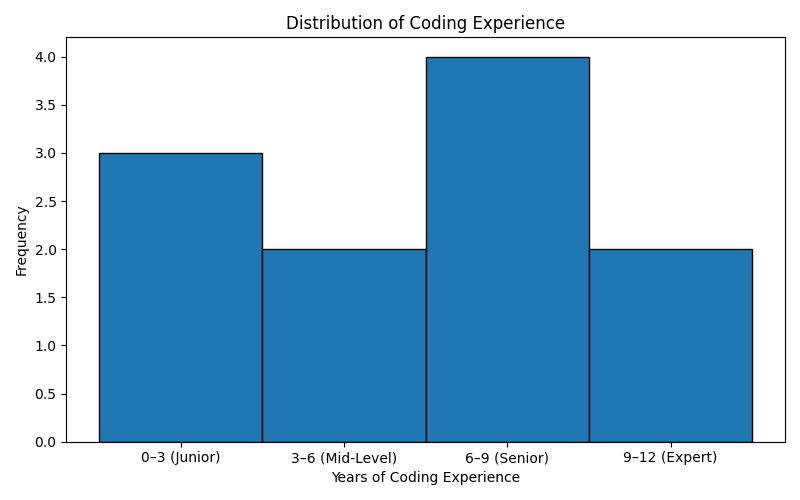}}
\caption{User Research Panel's Coding Experience (Years), Histogram}
\label{fig:yoe}
\end{figure}

To reflect the targeted user group, we recruited 11 participants with diverse professional backgrounds and coding experience.
The participants are recruited through the authors' networks and social media posts.
Our participants came from diverse professional sectors, including academia, finance, automotive, telecommunication and civil construction.
The organization sizes vary from small companies with 20-50 employees to large organizations with 300,000+ employees.
Their residence countries include The Netherlands, France, Germany, China and Japan.
All reported using IDEs and prompt-based LLM applications regularly in their work and had between 1 and 10 years of coding experience.
The distribution of participants' coding experience can be found from Figure~\ref{fig:yoe}.

\subsection{Data Collection}
We employed a usability-testing protocol suitable for both in-person and remote studies~\cite{boswell_alexander_moderated_2025}.  
The plugin and JetBrains IntelliJ IDEA IDE were run on the surveyor’s laptop, with offline participants interacting directly and online participants gaining remote control via Zoom screen-sharing.  
At the beginning of the study, the moderator briefed each participant on the plugin, the goal of the study, and the tasks to be performed.
With consent for data collection obtained, the moderator also encouraged the participants to think aloud to capture real-time reasoning.  
Throughout the tasks the moderator remained present with minimal intervention while still providing clarifications when requested.  
Immediately after task completion, participants filled out the SUS and raw NASA Task Load Index (NASA-TLX) survey on Google Forms.

\section{Results}
\subsection{RQ1: LLM Prompt Categorisation}

\begin{table}[tb]
\centering
\caption{Fleiss' Kappa Inter-Rater Agreement for OLD and NEW Prompt Categories}
\begin{tabular}{lcc}
\toprule
\textbf{Category} & \textbf{Fleiss' Kappa} & \textbf{Agreement Level~\cite{landis_measurement_1977}} \\
\midrule
OLD SDLC & 0.4315 & Moderate \\
OLD ROLE & 0.6898 & Substantial \\
OLD INTENT & 0.6253 & Substantial \\
OLD TYPE & 0.5573 & Moderate \\
\textbf{TOTAL (OLD Categories)} & \textbf{0.7219} & \textbf{Substantial} \\
\midrule
NEW SDLC & 0.4433 & Moderate \\
NEW ROLE & 0.6909 & Substantial \\
NEW INTENT & 0.6405 & Substantial \\
NEW TYPE & 0.5377 & Moderate \\
\textbf{TOTAL (NEW Categories)} & \textbf{0.7237} & \textbf{Substantial}
\\
\bottomrule 
\end{tabular}
\label{tab:complete_annotation_kappa}
\end{table}

Inter-rater agreement among the four LLM annotators for prompt categorization was substantial for both old and new prompts (overall \(\kappa=0.72\); Table~\ref{tab:complete_annotation_kappa}). 
Among the individual categories, Role labels achieved the highest consistency (\(\kappa=\) 0.69  for both old and new prompts), closely followed by Intent (\(\kappa=\) 0.63 and 0.64). 
Agreement on Prompt Type was moderate (\(\kappa=\) 0.56 and 0.54), while SDLC Phase exhibited the lowest yet still moderate reliability (\(\kappa=\) 0.43 and 0.44).
We find that the annotation approach is less effective on SDLC phase, as Claude-3-Haiku heavily favors "Planning \& Design", DeepSeek-Chat and OpenAI favors "General", each annotated more than half of all promps in their respective favored category, which Mistral showed a more balanced distribution.
According to the interpretation scale proposed by Landis and Koch~\cite{landis_measurement_1977}, these results suggest that current LLMs can classify software engineering prompts with moderate to substantial consistency.

\subsection{RQ2: Usability and User Experience}

\begin{figure}[tb]
\centerline{\includegraphics[width=0.85\columnwidth]{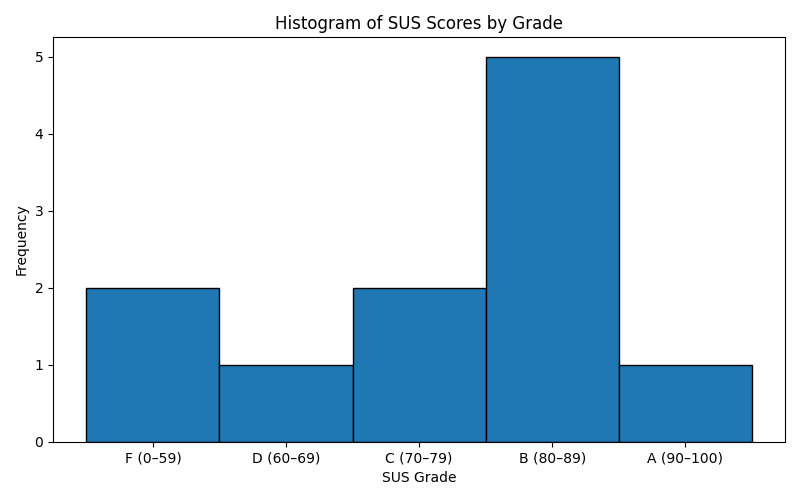}}
\caption{Histogram of overall SUS Scores by Approximate Grading~\cite{james_r_jim_lewis_phd_item_2018}}
\label{fig: sus}
\end{figure}

\begin{figure}[tb]
\centerline{\includegraphics[width=0.9\columnwidth]{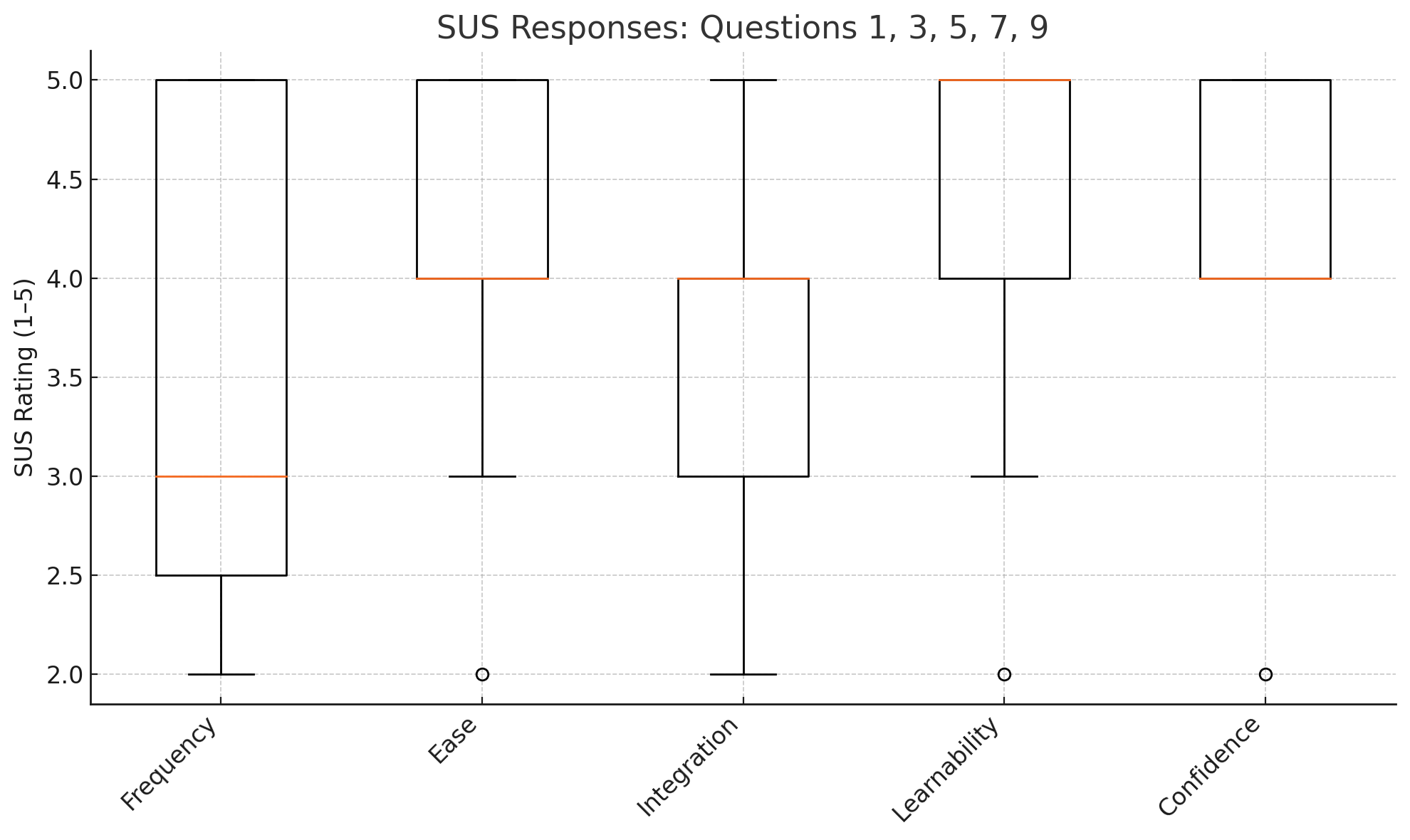}}
\caption{Score Distribution for SUS question 1, 3, 5, 7, 9 (Positive Usability)}
\label{fig:sus_box_good}
\end{figure}

\begin{figure}[tb]
\centerline{\includegraphics[width=0.9\columnwidth]{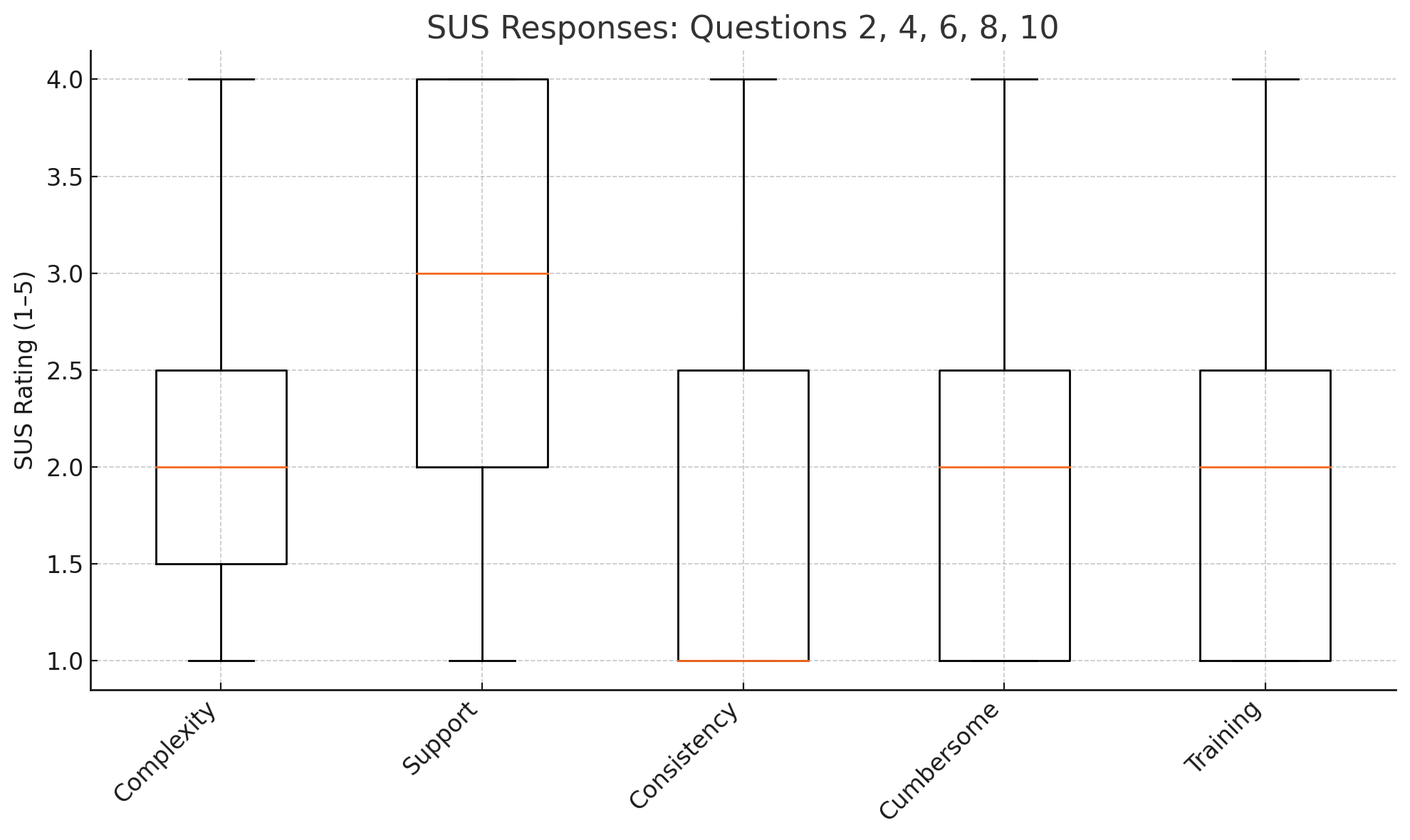}}
\caption{Score Distribution for SUS question 2, 4, 6, 8, 10 (Negative Usability)}
\label{fig:sus_box_bad}
\end{figure}

Participants generally perceived the system as highly usable, with a mean SUS score of 72.73 out of 100 [CI 95\% (61.54, 83.91)]. According to the interpretation guidelines proposed by Lewis~\cite{james_r_jim_lewis_phd_item_2018}, this score corresponds to a ``good'' user experience, suggesting that the system meets the usability expectations of its target audience.
Figure~\ref{fig:sus_box_good} presents the participants’ ratings across the evaluated usability aspects, where higher values indicate better perceived usability. Overall, the system received strong ratings for ease of use, learnability, and user confidence, with nearly all participants scoring these aspects highly. Integration also received favorable evaluations; however, it exhibits the widest distribution of ratings, suggesting that while the components work together reasonably well, there is noticeable room for improvement in their integration. The most divergent opinions concern the expected frequency of tool usage, with responses spanning the full 2–5 range and clustering around a median of 3. This pattern reflects a mixed perception: some developers envision the tool as part of their daily workflow, whereas others anticipate using it only occasionally. Notably, no participant indicated that they would completely avoid using the tool. This highlights at least some perceived utility for all users.

Figure~\ref{fig:sus_box_bad} presents participants’ ratings for each usability question, where higher values indicate poorer usability. Four items, complexity, consistency, cumbersome usage, and training requirements, show median ratings of 2 or lower, suggesting that most participants disagreed with statements implying that Prompt-with-Me is overly complex, inconsistent, awkward, or difficult to learn. The only notable exception is the ``support'' item, which reflects mixed opinions regarding the need for additional guidance. Across all items, boxplots span approximately two scale points, with whiskers reaching both extremes, indicating some variability among responses. Overall, these results suggest that while the majority of participants experienced smooth and intuitive interaction, a small subset still perceived notable usability challenges.

In general, the distribution suggests generally positive usability, with some variability in anticipated usage frequency and an indication that extra support may be required.

\subsection{RQ3: Helpfulness of Prompt-with-Me}

We assessed helpfulness through participants’ open-ended survey responses and reflections during the study tasks.
All eleven participants reported that Prompt-with-Me accelerated their prompt-writing process, primarily due to its \textbf{template generation} feature.

Seven participants highlighted that it eliminated the need to copy, paste, or re-type similar prompts repeatedly.
As Participant \#10 explained, ``\textit{A lot of my prompts are almost the same, so typing them out each time is tedious. The system let me pick from my past prompts and reminded me of the exact wording I used before. It saved me a ton of time.}''

Other features, such as \textbf{automatic classification, anonymization, grammar correction, and similarity checking} were also noted as time-savers.
Several participants estimated that the integrated IDE plugin saved ``\textit{a few minutes per prompt}'', which accumulates over a full day of coding.
For example, Participant \#6 shared, ``\textit{When I am scripting in Revit’s Dynamo, I hit the same scenarios across different maquettes. Being able to pull up and reuse the right prompts speeds up programming across all my 3D models.}''

Overall, participants agreed that the current version of Prompt-with-Me not only reduces repetitive work but also improves the quality of prompts.

\subsection{RQ4: Perceived Cognitive Load}

\begin{figure}[tb]
\centerline{\includegraphics[width=0.9\columnwidth]{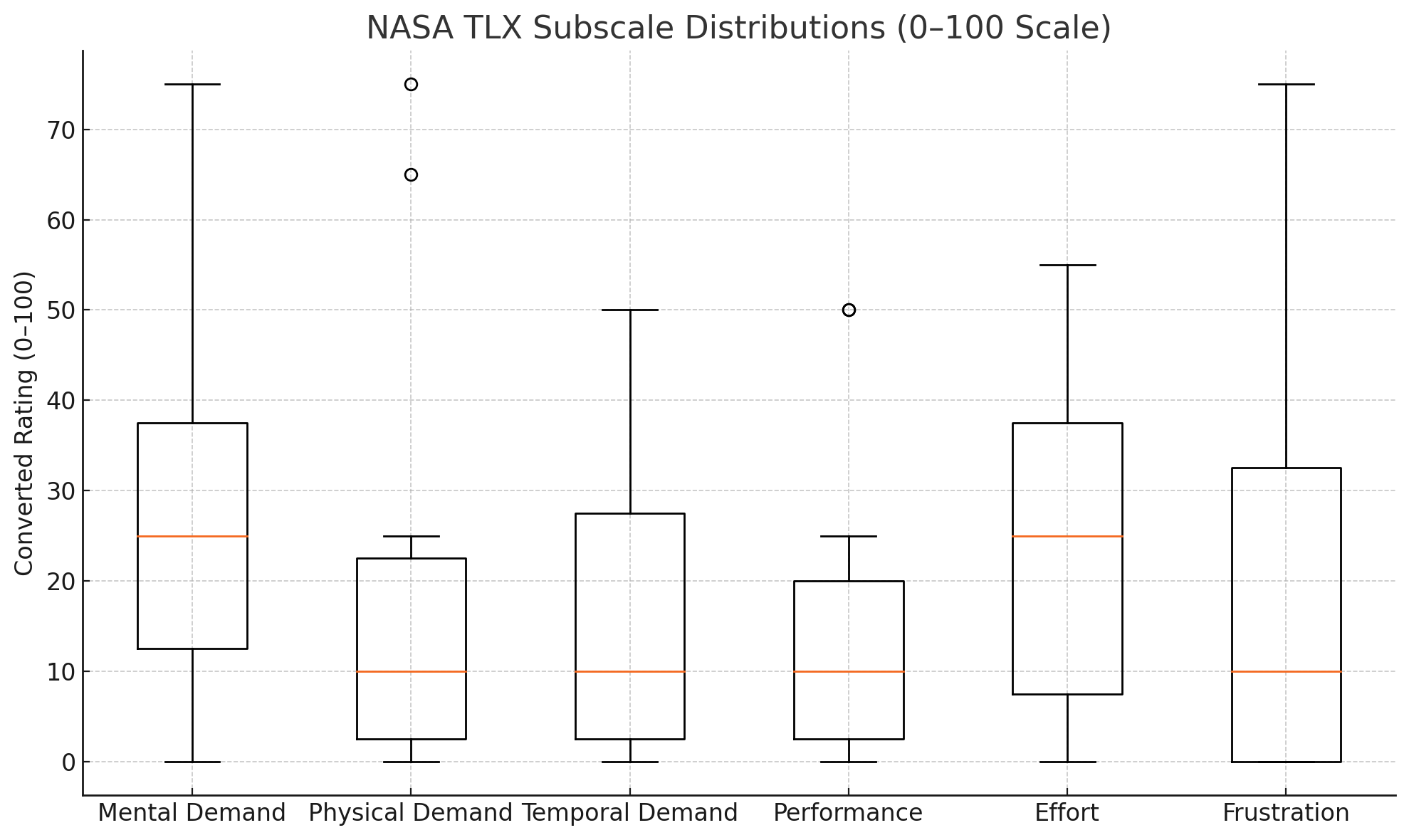}}
\caption{NASA TLX Subscale Distributions}
\label{fig:tlx_box}
\end{figure}

\begin{table}[tb]
\centering
\caption{Raw TLX Score reported by User Study Participants}
\begin{tabular}{lcc}
\hline
Dimension        & Mean (0–100) & 95 \% CI \\ 
\hline
Mental demand    & 29.55        & [14.55, 44.54] \\ 
Physical demand  & 20.45        & [3.98, 36.93]  \\ 
Temporal demand  & 17.27        & [5.02, 29.52]  \\ 
Performance      & 15.91        & [4.21, 27.61]  \\ 
Effort           & 23.64        & [10.92, 36.35] \\ 
Frustration      & 19.55        & [3.38, 35.71]  \\ 
\hline
\textbf{Overall} & \textbf{21.06}& \textbf{[9.90, 32.23]} \\
\hline
\end{tabular}
\label{tab:tlx_scores}
\end{table}

\begin{table}[tb]
  \centering
    \caption{Interpretation bands for overall NASA TLX scores (0–100)~\cite{brillinger_physiological_2024}}%
  \begin{tabular}{@{}ll@{}}
    \toprule
    \textbf{TLX Score (0–100)} & \textbf{Interpretation} \\
    \midrule
    0–9       & Low workload         \\
    10–29     & Medium workload      \\
    30–49     & Somewhat high workload \\
    50–79     & High workload        \\
    80–100    & Very high workload   \\
    \bottomrule
  \end{tabular}
  \label{tab:tlx-interpretation}
\end{table}

Figure~\ref{fig:tlx_box} presents the six raw NASA-TLX subscale scores on a 0–100 scale, and Table~\ref{tab:tlx_scores} reports their means and 95\% confidence intervals. Across all subscales, the mean scores fall within the medium workload range (10–29) defined in Table~\ref{tab:tlx-interpretation}. Mental demand was the highest reported factor [Mean=29.55, CI 95\% (14.55, 44.54)], approaching the threshold for ``somewhat high'' workload, whereas all other subscales remained solidly in the medium range. The overall workload score was 21.06, [CI 95\% (14.55, 44.54)]. This ensures that Prompt-with-Me imposed only a \textbf{light cognitive load} on participants.

The boxplots in Figure~\ref{fig:tlx_box} reveal long whiskers and several outliers, particularly for mental demand, physical demand, and frustration, suggesting that a small number of participants experienced brief periods of higher cognitive load. Nevertheless, the overall distribution patterns confirm that interacting with Prompt-with-Me generally imposed only \textbf{modest cognitive effort}.

\section{Discussion}

\subsection{Insights from the Controlled Evaluation}
\textbf{Template Extraction Enables Prompt Reuse}:
Prompt-with-Me consolidates similar prompts into a single, parameterized template to enable systematic reuse and adaptation. These templates function as human-readable programs: they combine a stable core corpus with variable components that can be adjusted to different application scenarios. In our user study, a majority of participants (7 out of 11) identified template generation as the most valuable feature of Prompt-with-Me. Beyond improving efficiency, templating supports versioning and reviewing prompts as stable, reusable artifacts within a software project. This approach also has the potential to reduce errors commonly introduced through ad hoc copy-and-paste prompt editing~\cite{liang_prompts_2025}, resulting in both consistency and reliability in LLM-based workflows.

\textbf{Tool Adaptiveness Drives Adoption More Than Low Cognitive Load}:
Although participants consistently reported a low cognitive load when using the tool, their anticipated frequency of use varied considerably. This suggests that ease of use alone does not guarantee frequent adoption. Participants from various sectors emphasized that perceived usefulness depends on how well Prompt-with-Me adapts to the specific workflows and practices of their domain. Consequently, the ability of the tool to integrate into established routines and support customization of existing workflows emerges as a critical factor for sustained adoption and a worthy avenue for future work.

\subsection{Lessons Learned for Future Designs}
Our study revealed several design lessons that tool builders should consider to enhance usability and user trust. 

\textbf{Transparency and Reversibility to Enhance User Trust}:
First, participants consistently emphasized that automation is only welcomed when its effects are both transparent and reversible. Users want to understand what the tool has changed and feel confident that they can easily undo or adjust these modifications. This finding highlights the importance of providing clear, accessible explanations of any optimizations or transformations performed by the tool, along with straightforward mechanisms for reverting them. 

\textbf{Need for Domain-Specific Customization in Taxonomy Design}: 
Although participants from diverse backgrounds generally accepted the four-dimensional taxonomy, users from more specialized domains expressed a need for a more customizable taxonomy that aligns closely with their specific workflows. This highlights the importance of developing extendable classification engines capable of accommodating domain-specific requirements to ensure that the system can adapt to the nuanced practices of different user groups.

\textbf{Simplify UI}:
Some participants noted that the main window of the all-in-one plugin could benefit from a simpler, more streamlined design. While the current interface effectively presents all the information required for prompt optimization, it also contributes to an already crowded IDE environment. Beyond functional improvements, participants suggested that greater attention should be paid to the visual and interaction design to enhance usability and reduce cognitive load.

\textbf{Balance Privacy and Performance}:
Participants from privacy-sensitive organizations preferred \textit{local} processing for all prompt optimizations, while others prioritized \textit{speed} and \textit{performance} over data locality. This contrast suggests that providing a flexible option, such as a toggle between online and offline modes, could accommodate both needs and potentially broaden adoption.

\textbf{Context-Aware Optimizations Increase Value}:
Beyond deployment preferences, participants noted that prompt optimizations could be enhanced by leveraging \textit{IDE context} and project-specific information.

\textbf{Collaboration Features Enable Teamwide Benefits}:
Several participants highlighted the need for \textit{collaborative features}. Suggested capabilities included a shared prompt library with version control, in-line comments, and review mechanisms, enabling teams to build and refine prompts collectively.

\subsection{Threats to the Validity}

\textbf{Internal validity}:
Our user study involved eleven participants, which, while sufficient for a qualitative and exploratory evaluation, may not fully capture the range of developer experiences. Additionally, usefulness and mental effort were assessed through self‑reported questionnaires. Although subjective measures are a standard first step in evaluating developer tools, future studies can complement them with objective metrics such as task completion time, TLX scores, or usage logs to strengthen reliability.

\textbf{External validity}: 
Our study focused on SQL‑based tasks within JetBrains IntelliJ using controlled scenarios rather than live production environments. While this allowed us to ensure consistency and isolate the effects of Prompt‑with‑Me, it naturally narrows generalizability to other IDEs, domains, or workflow contexts. Expanding future evaluations to larger and more diverse participant pools will help confirm and broaden our conclusions.

\textbf{Construct validity}: 
We operationalized usefulness and mental effort through questionnaire responses and inferred template reuse from self‑reports rather than direct behavioral or long‑term adoption data. While sufficient for a first investigation, future work can strengthen construct validity by combining subjective responses with behavioral logs, longitudinal studies, and productivity indicators.

\section{Conclusion}
Prompts are now critical in LLM-enabled software development, yet they are often crafted and managed in an ad hoc manner. Prompt-with-Me addresses this gap by demonstrating how prompts can be systematically organized, maintained, and leveraged within development workflows. Our taxonomy study reveals that modern LLMs can reliably classify software engineering prompts, providing a foundation for structured prompt management. Complementing this, our user study shows that developers appreciate Prompt-with-Me for its usability, efficiency, and low cognitive overhead. Looking ahead, we plan to enrich the taxonomy, introduce collaborative features, and refine the plugin’s interface. By integrating prompt maintenance into everyday programming tools, Prompt-with-Me takes a meaningful step toward safer, more sustainable, and developer-friendly LLM-based software development.

\bibliographystyle{ieeetr}
\bibliography{main}

\end{document}

%% file: tables/annotation_results.tex
\begin{table}[tb]
\centering
\caption{Single-Model Contribution to Fleiss' \(\kappa\) of All Models (\(K_4 -K_{3,\overline{i}}\))}
\begin{tabular}{|l|c|c|c|c|}
\hline
\textbf{Category} & \textbf{Haiku} & \textbf{DS} & \textbf{Mistral} & \textbf{GPT-4o} \\
\hline
OLD, SDLC& -0.0316 & -0.0061 & \textbf{+0.0359} & -0.0378 \\
OLD, ROLE& -0.0316 & \textbf{+0.0255} & +0.0240 & -0.0145 \\
OLD, INTENT& -0.0031 & -0.0048 & \textbf{+0.0240} & -0.0145 \\
OLD, TYPE& +0.0325 & -0.0203 & \textbf{+0.0449} & -0.0519 \\
NEW, SDLC& -0.0316 & -0.0061 & \textbf{+0.0360} & -0.0378 \\
NEW, ROLE& -0.0316 & \textbf{+0.0255} & +0.0240 & -0.0145 \\
NEW, INTENT& -0.0031 & -0.0048 & \textbf{+0.0240} & -0.0145 \\
NEW, TYPE& +0.0325 & -0.0203 & \textbf{+0.0449} & -0.0519 \\
\hline
\textbf{Category Wins} & 0 & 2 & \textbf{6} & 0 \\
\hline
\end{tabular}
\label{tab:leave_one_out_fleiss}
\small
\end{table}

\begin{table}[tb]
\caption{Mistral Dataset Distribution}
\centering
\begin{tabular}{|l|c|c|}
\hline
\textbf{Category} & \textbf{Old Prompts} & \textbf{New Prompts}\\
\hline
\hline
\multicolumn{3}{|c|}{\textbf{SDLC Phase}} \\
\hline
General & 446 & 444 \\
Implementation \& Coding & 279 & 284 \\
Testing \& Quality Assurance & 213 & 213 \\
Planning \& Design & 170 & 167 \\
\hline

\multicolumn{3}{|c|}{\textbf{Role}} \\
\hline
General & 686 & 685 \\
Software Developer & 318 & 316 \\
Project Manager & 68 & 67 \\
Data Scientist & 36 & 40 \\
\hline

\multicolumn{3}{|c|}{\textbf{Intent}} \\
\hline
Best Practices & 513 & 517 \\
Documentation \& Explanation & 269 & 262 \\
Code Generation & 262 & 266 \\
Code Review \& Analysis & 64 & 63 \\
\hline

\multicolumn{3}{|c|}{\textbf{Type}} \\
\hline
Template-based & 748 & 768 \\
Zero-shot & 205 & 180 \\
Few-shot & 155 & 160 \\
\hline
\end{tabular}
\label{tab:mistral_comparison}
\end{table}